\documentclass[preprint,showpacs,preprintnumbers,amsmath,amssymb,prb]{revtex4-1}

\usepackage[dvipdfmx]{graphicx}
\usepackage{color}
\usepackage{bm}% bold math
\usepackage{colortbl}
\usepackage{calc}
\usepackage{multirow,bigdelim}
\usepackage{comment}
\usepackage{accents}
\usepackage{braket}
\usepackage{mathdots}
\usepackage{datetime2}
\usepackage{ulem}

\newcommand{\mat}[1]{\mathbf{#1}}
\begin{document}

% Use the \preprint command to place your local institutional report
% number in the upper righthand corner of the title page in preprint mode.
% Multiple \preprint commands are allowed.
% Use the 'preprintnumbers' class option to override journal defaults
% to display numbers if necessary
%\preprint{Preprint (REV\TeX\ 4.2): \DTMnow}

%Title of paper
\title{Efficient calculation of the Green's function in scattering region for electron-transport simulations}

% repeat the \author .. \affiliation etc. as needed
% \email, \thanks, \homepage, \altaffiliation all apply to the current
% author. Explanatory text should go in the []'s, actual e-mail
% address or url should go in the {}'s for \email and \homepage.
% Please use the appropriate macro foreach each type of information

% \affiliation command applies to all authors since the last
% \affiliation command. The \affiliation command should follow the
% other information
% \affiliation can be followed by \email, \homepage, \thanks as well.

%%%
%%% Author 1
%%%
\author{Yoshiyuki Egami}
%\homepage[]{Your web page}
%\altaffiliation{}
%\thanks{These authors equally contributed to this work.}
\email[Corresponding author: ]{y\_egami@eng.hokudai.ac.jp}
\affiliation{Division of Applied Physics, Faculty of Engineering, Hokkaido University, Sapporo, Hokkaido 060-8628, Japan}

%%%
%%% Author 2
%%%
\author{Shigeru Tsukamoto}
%\email[Corresponding author: ]{s.tsukamoto@fz-juelich.de}
%\homepage[]{Your web page}
%\altaffiliation{}
%\thanks{These authors equally contributed to this work.}
%\email[Corresponding author: ]{s.tsukamoto@fz-juelich.de}
\affiliation{Peter Gr{\"u}nberg Institut \& Institute for Advanced Simulation, Forschungszentrum J{\"u}lich and JARA, D-52425 J{\"u}lich, Germany}

%%%
%%% Author 3
%%%
\author{Tomoya Ono}
%\thanks{These authors equally contributed to this work.}
%\email[Corresponding author: ]{t.ono@eedept.kobe-u.ac.jp}
\affiliation{Department of Electrical and Electronic Engineering, Kobe University, Kobe, Hyogo 657-8501, Japan}
%Collaboration name if desired (requires use of superscriptaddress
%option in \documentclass). \noaffiliation is required (may also be
%used with the \author command).
%\collaboration can be followed by \email, \homepage, \thanks as well.
%\collaboration{}
%\noaffiliation

\date{\today}

\begin{abstract}
We propose a first-principles method of efficiently evaluating electron-transport properties of very long systems. 
Implementing the recursive Green's function method and the shifted conjugate gradient method in the transport simulator based on real-space finite-difference formalism, we can suppress the increase in the computational cost, which is generally proportional to the cube of the system length to a linear order.
This enables us to perform the transport calculations of double-walled carbon nanotubes~(DWCNTs) with 196,608 atoms. We find that the conductance spectra exhibit different properties depending on the periodicity of doped impurities in DWCNTs and they differ from the properties for systems with less than 1,000 atoms.
\end{abstract}

% insert suggested keywords - APS authors don't need to do this

%\maketitle must follow title, authors, abstract, \pacs, and \keywords
\maketitle

%%%
%%% %\section{Introduction\label{sec:Introduction}}
%%%
One-dimensional materials such as nanowires and nanotubes, which have unique electronic properties due to the quantum confinement effect, are expected to be applied to electronic and spintronic devices, optoelectronic circuits, and biosensors.\cite{Science287-000622,Nature409-000066,AdvMater14-000158,LI200618,AC078-004260,NanoLett6-001087,PRB75-075410} In recent years, large-scale electron-transport calculations have been indispensable for designing the functionality of electronic devices.
Although first-principles calculations based on the density functional theory~(DFT)\cite{RevModPhys71-001253,*PhysRev136-00B864,*PhysRev140-0A1133} allow us to accurately evaluate electron-transport properties of atomic systems, their target is typically limited to small systems because of the heavy computational cost arising from calculating the Green's functions, which is intrinsic cubic scaling with system size. 
Limited-scale system is sufficient to simulate simple systems, such as locally perturbed bulk regions or interfaces. However, it is not suitable for simulating complex systems, such as bulk with a realistic defect density, amorphous structure, and interfaces with lattice mismatches between different crystalline materials.
Thus far, to circumvent this restriction, transport calculations for large systems containing several thousands of atoms have been performed within atomic-basis formalism and tight-binding (TB) formalism based on DFT.\cite{PRB96-161404R,SISPAD2019-8870571,Nanoscale11-006153,JAP120-154301} Since the Hamiltonian matrix is dense in these formalisms, the computational cost of the inverse matrix for obtaining the Green's function matrix by the direct method is proportional to the cube of the matrix dimension.
Moreover, there are unavoidable problems such as the incompleteness of basis sets and inefficiency in parallel computing.
However, real-space finite-difference (RSFD) formalism is also recognized as suitable for large-scale calculations requiring high computational accuracy because it has a high affinity to massively parallel architectures and can avoid problems arising from the basis sets.\cite{PRB67-195315,KikujiHirose2005,PRB86-195406,PRE91-063305,PRE92-033301,PRB93-045421,PRE95-033309,*PRB97-115450,*PRB98-195422,PRB100-075413}

\begin{figure}
    \includegraphics{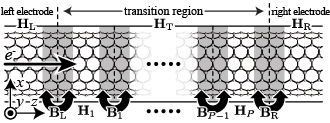}
    \caption{\label{fig:Figure1} Schematic representation of a junction system. The system is partitioned into one transition region and left and right semi-infinite electrodes, whose Hamiltonians are denoted by $\mat{H}_{\mathrm{T}}$, $\mat{H}_{\mathrm{L}}$, and $\mat{H}_{\mathrm{R}}$, respectively. $\mat{B}_{\mathrm{L(R)}}$ denotes the interaction between adjoining small parts. The transition region consists of $P$ parts. The Hamiltonians are denoted by $\mat{H}_{i}$ for $i=1,\dots,P$, and the interactions between the adjoining parts by $\mat{B}_{i}$ for $i=1,\dots,P-1$. }
\end{figure}

Fujimoto and Hirose developed the overbridging boundary matching (OBM) method based on the RSFD formalism\cite{PRB67-195315} by exploiting the advantages for electron-transport calculations, where a whole system is divided along the $z$ direction (see Fig.~\ref{fig:Figure1}) into three regions: the left electrode, the transition region and the right electrode.
The electron-transport calculation refers only to subsets of the Green's function matrix of the transition region.
In the OBM method, the subsets are calculated efficiently using a shifted conjugate-gradient (SCG) solver\cite{PRE91-063305} since the Hamiltonian matrix in the RSFD formalism is sparse.
In the SCG solver, the computational cost for calculating the subsets is ${\cal{O}}(N_{E}N_{xy}^{3}N_{z})$ when the algorithm proposed by Takayama {\it{et al.}}\cite{PRB73-165108} is adopted, where $N_{xy}$ and $N_{z}$ are the numbers of grid points in the $xy$ plane and the $z$ direction, respectively, and $N_E$ denotes the number of energy values $E$ to be treated at once by the SCG method.
For systems with a large $N_z$, the computational cost for matrix-vector operations, which is ${\cal O}(N_{xy}^{3}N_{z}^2)$, becomes dominant.\footnote{Although the computational cost of the subsets of the Green's function matrix is generally dominant, the cost of coefficient matrix-vector operations dominates in the case that $E$ becomes close to the eigenvalue of the truncated Hamiltonian matrix. The matrix-vector operation cost also becomes dominant when $N_z$ is large because the density of the eigenvalues in energy increases.} Although the maximum order of the computational cost is reduced from ${\cal O}(N_{xy}^{3}N_{z}^3)$ to ${\cal O}(N_{xy}^{3}N_{z}^2)$ by the OBM method,
the transport calculation has room for further reducing the computational cost to handle more realistic and longer one-dimensional systems.

%%%
%%% Figure1
%%%

In this letter, we propose an efficient computational procedure based on RSFD formalism to evaluate the electron-transport properties of long systems containing more than 100,000 atoms without accuracy deterioration.
We verify the computational accuracy and performance of the proposed method.
Additionally, to exemplify the efficiency of the proposed procedure we demonstrate large-scale electron-transport calculations for double-walled carbon nanotubes (DWCNTs) composed of 196,608 atoms, which make, to the best of our knowledge, the largest system in the first-principles electron-transport calculation.

%%%%%%%%
%%%%%%%% computational scheme
%%%%%%%%
Thouless and Kirkpatrick proposed an efficient calculation of subsets of a Green's function matrix for a transition region within TB formalism as the recursive Green's function (RGF) method.\cite{JPhysC14-000235}
The RGF method constructs a transition region by arranging multiple parts along the $z$ direction as illustrated in Fig.~\ref{fig:Figure1}, and calculates only necessary subsets by recursively combining the Green's function matrices of the adjoining small parts one by one.
Since the calculation of the Green's function matrices for the small parts is less costly than that for the extended transition region, the RGF method is capable of handling large systems and is widely used for calculating the transport properties of atomic-basis models and TB models.\cite{JComputElectron12-000203,*JComputPhys215-000741,pp174ch3DavidKFerry1997,FernandoSols1995,GeorgoMetalidis2007,AIPAdvances2-032115,JAP119-154505,PhysChemChemPhys21-026027}

Now, we discuss the advantage of the RGF method in transport calculations within RSFD formalism. The transport properties are evaluated using the Green's function of the transition region suspended between semi-infinite electrodes, $\mat{G}_{\mathrm{T}}$, which is constructed by the Green's function associated with the truncated transition region, $\mat{\cal G}_{\mathrm{T}}$, and the self-energy terms of the electrodes.\cite{PRB86-195406} The Green's function matrix $\mat{\cal G}_{\mathrm{T}}=(E\mat{I}_{\mathrm{T}}-\mat{H}_{\mathrm{T}})^{-1}$, in which $\mat{H}_{\mathrm{T}}$ is the truncated Hamiltonian matrix of the extended transition region consisting of $P$ parts (see Fig.~\ref{fig:Figure1}), is expressed within the framework of RSFD formalism using the norm-conserving pseudopotentials~(NCPPs) as
{%
\renewcommand{\arraystretch}{0}
\newlength\celldim
\newlength\fontheight
\newlength\fontdepth
\newlength\extraheight
\settoheight{\fontheight}{\mathstrut}
\settodepth{\fontdepth}{\mathstrut}
\setlength{\celldim}{2\fontheight}
\setlength{\extraheight}{\celldim-\fontheight-\fontdepth}
\newcolumntype{S}{@{}>{\centering\arraybackslash}p{\celldim}<{\rule[-0.5\extraheight]{0pt}{\celldim}}@{}}
%\newcolumntype{C}{@{}>{\centering\arraybackslash}c<{\rule[-0.5\extraheight]{0pt}{\celldim}}@{}}
\newcolumntype{C}{>{\centering\arraybackslash}c<{\rule[-0.5\extraheight]{0pt}{\celldim}}}
%celldim = \the\celldim, extraheight = \the\extraheight, fontheight = \the\fontheight, fontdepth = \the\fontdepth
\begin{equation}
\mat{\cal G}_{\mathrm{T}}=\left(E\mat{I}_{\mathrm{T}}-\left[
\begin{array}{SSSSSSS}
  \cline{1-2}
  \multicolumn{1}{|S}{} & \multicolumn{1}{S|}{} & & & & & \\
  \cline{3-3}
  \multicolumn{2}{|c|}{\raisebox{0.5\celldim}{\smash{$\mat{H}_{1}$}}} & \multicolumn{1}{S|}{$\mat{B}_{1}$} & & & \multicolumn{2}{c}{\raisebox{0.5\celldim}{\smash{$\mat{0}$}}} \\
  \cline{1-4}
  \multicolumn{1}{S|}{} & \multicolumn{1}{c|}{\mat{B}_{1}^{\dag}} & & \multicolumn{1}{S|}{} & \\
  \cline{2-2}\cline{5-5}
  & \multicolumn{1}{c|}{} & \multicolumn{2}{c|}{\raisebox{0.5\celldim}{\smash{$\mat{H}_{2}$}}} & \multicolumn{1}{S|}{$\mat{B}_{2}$} & \\
  \cline{3-5}
  & & \multicolumn{1}{S|}{} & \multicolumn{1}{S|}{$\mat{B}_{2}^{\dag}$} & $\ddots$ & $\ddots$ & \\
  \cline{4-4}\cline{6-7}
  & & & & \multicolumn{1}{S|}{\smash{$\ddots$}} & & \multicolumn{1}{S|}{} \\
  \multicolumn{2}{c}{\raisebox{0.5\celldim}{\smash{$\mat{0}$}}} & & & \multicolumn{1}{S|}{} & \multicolumn{2}{c|}{\raisebox{0.5\celldim}{\smash{$\mat{H}_{P}$}}} \\
  \cline{6-7}
\end{array}\right]\right)^{-1},
\label{eq:GreenFunction}
\end{equation}
}%
where $\mat{H}_i$ is the truncated Hamiltonian matrix for the transition region of the $i$th part.
For simplicity, we assume that all $\mat{H}_i$ ($i=1,\ldots,P)$ have the same dimension, the interactions between the second- and further-neighboring parts are zero, and the non-zero matrix, $\mat{B}_{i}$, represents the interaction between the adjoining parts. 
We define the dimensions of $\mat{B}_{i}$ and $\mat{H}_i$ as $M_{i}\times N_{i}$ and $N_{xy}N_z^{(\text{part})}$, respectively, with $N_z^{(\text{part})}=N_z/P$; $\mat{I}_{\mathrm{T}}$ denotes an $N_{xy}N_{z}$-dimensional identity matrix.

Since the Hamiltonian matrix of RSFD formalism is highly sparse, it is more advantageous to calculate the subsets of Green's function matrix of the transition region by iterative solvers than it is to compute the entire Green's function by direct inversions of the Hamiltonian matrix.
According to Ref.~\onlinecite{PRB86-195406}, the electron-transport properties can be estimated using submatrices located at the four corners of the Green's function matrix, $\mat{\cal G}_{\mathrm{T}}$. 
Now we will combine the two adjoining parts to derive the Green's function matrix associated with a truncated Hamiltonian composed of the $i$th and ($i+1$)th parts, which is denoted by the $2N_{xy}N_z^{(\text{part})}$-dimensional matrix $\mat{\cal G}_{i:i+1}$.
As described in Supplemental Material, the submatrices located at the upper-left, upper-right, lower-left, and lower-right corners of $\mat{\cal G}_{i:i+1}$ are given as
\begin{eqnarray}
 & &
\begin{split}
\mat{\cal G}_{i:i+1}^{\mathrm{UL}} & \!=\! \mat{\cal G}_{i:i}^{\mathrm{UL}}+\mat{\cal G}_{i:i}^{\mathrm{UR}}\mat{B}_{i}\mat{\cal G}_{i+1:i+1}^{\mathrm{UL}}\mat{B}_{i}^{\dag} \\
                                   & \quad\times[\mat{I}_{M_{i}}-\mat{\cal G}_{i:i}^{\mathrm{LR}}\mat{B}_{i}\mat{\cal G}_{i+1:i+1}^{\mathrm{UL}}\mat{B}_{i}^{\dag}]^{-1}\mat{\cal G}_{i:i}^{\mathrm{LL}},
\end{split}
\label{eq:GTUL} \\
 & &
\begin{split}
\mat{\cal G}_{i:i+1}^{\mathrm{UR}} & \!=\! -\mat{\cal G}_{i:i}^{\text{UR}}[\mat{I}_{M_{i}}\!-\!\mat{B}_{i}\mat{\cal G}_{i+1:i+1}^{\mathrm{UL}}\mat{B}_{i}^{\dag}\mat{\cal G}_{i:i}^{\mathrm{LR}}]^{-1} \mat{B}_{i}\mat{\cal G}_{i+1:i+1}^{\mathrm{UR}}, 
\end{split}
\label{eq:GTUR} \\
 & &
\begin{split}
\mat{\cal G}_{i:i+1}^{\mathrm{LL}} & \!=\! -\mat{\cal G}_{i+1:i+1}^{\text{LL}}[\mat{I}_{N_{i}}\!-\!\mat{B}_{i}^{\dag}\mat{\cal G}_{i:i}^{\mathrm{LR}}\mat{B}_{i}\mat{\cal G}_{i+1:i+1}^{\mathrm{UL}}]^{-1} \mat{B}_{i}^{\dag}\mat{\cal G}_{i:i}^{\mathrm{LL}}, 
\end{split}
\label{eq:GTLL} \\
 & &
\begin{split}%
\mat{\cal G}_{i:i+1}^{\mathrm{LR}} & \!=\! \mat{\cal G}_{i+1:i+1}^{\mathrm{LR}}+\mat{\cal G}_{i+1:i+1}^{\mathrm{LL}}\mat{B}_{i}^{\dag}\mat{\cal G}_{i:i}^{\mathrm{LR}}\mat{B}_{i} \\
                                   & \quad   \times[\mat{I}_{N_{i}}-\mat{\cal G}_{i+1:i+1}^{\mathrm{UL}}\mat{B}_{i}^{\dag}\mat{\cal G}_{i:i}^{\mathrm{LR}}\mat{B}_{i}]^{-1}\mat{\cal G}_{i+1:i+1}^{\mathrm{UR}}, 
\end{split}%
\label{eq:GTLR}
\end{eqnarray}%
respectively. 
$\mat{I}_{M_{i}(N_{i})}$ denotes an $M_{i}(N_{i})$-dimensional identity matrix, and $\mat{\cal G}_{i:i}$ represents the $N_{xy}N_z^{(\text{part})}$-dimensional Green's function matrix associated with $\mat{H}_i$, that is, $\mat{\cal G}_{i:i}=(E\mat{I}_i-\mat{H}_{i})^{-1}$, with $\mat{I}_i$ being an $N_{xy} N_z^{(\text{part})}$-dimensional identity matrix. 
As shown in Eqs.~\eqref{eq:GTUL}--\eqref{eq:GTLR}, we only need the four corners of $\mat{\cal G}_{i:i}$, which are efficiently obtained by using the SCG method.
In most cases, $M_i=N_i=N_{\text{f}}N_{xy}$, with $N_{\text{f}}$ being the order of the finite-difference approximation.\cite{PRB50-011355,*PRL72-001240}
Repeating this procedure to connect the parts numbered from $i$ though to $j$, we can construct the four submatrices of a $(j-i+1)N_{xy}N_z^{(\text{part})}$-dimensional matrix $\mat{\cal G}_{i:j}$. 

Now, we estimate the computational cost of the proposed method. 
We assume that $\mat{B}_{\mathrm{L(R)}}$ (see Fig.~\ref{fig:Figure1}) and $\mat{B}_{i}$ for $i=1,\dots,P-1$ are $M$-dimensional square matrices with $M=N_{\text{f}}N_{xy}$. 
In the case of the RGF method based on the atomic-basis formalism, $\mat{H}_{\mathrm{T}}$ is dense; that is, $M$ is not sufficiently smaller than $N_{xy}N_z^{(\text{part})}$. Therefore, the iterative solvers are not more efficient than direct inversion, and there is not much benefit in using Eqs.~\eqref{eq:GTUL}--\eqref{eq:GTLR} to reduce the computational cost.
Furthermore, the overlap matrix is generally not any scalar matrix; hence, the SCG method cannot be used to solve the linear systems.
However, the RGF algorithm must be compatible with RSFD formalism since $\mat{H}_{i}$ is a block-diagonal matrix and a sparse matrix.\cite{KikujiHirose2005,PRB50-011355,PRL72-001240} 
The proposed method can be implemented without negatively affecting the benefit of the SCG method, in which the matrix-vector operations for shifted energy points are omitted, and only the matrix elements of subsets are updated for shifted energy points during SCG iteration.
The computational cost of the matrix-vector products with a sparse matrix increases linearly with $N_z$, and the number of SCG iterations is proportional to $N_z$.
Therefore, the computational cost for the extended transition region is reduced from ${\cal O}(N_{xy}^{3}N_{z}^2\bigr)={\cal O}(N_{xy}^{3}(N_{z}^{\text{(part)}}P)^2\bigr)$ to ${\cal O}(N_{xy}^{3}(N_{z}^{\text{(part)}})^2P\bigr)$ using the RGF and the SCG methods since the cost is ${\cal O}(N_{xy}^{3}(N_{z}^{\text{(part)}})^2\bigr)$ for each of the $P$ parts; that is, the increase in computational cost can be suppressed linearly for $P$.
Although the computational cost for calculating Eqs.~\eqref{eq:GTUL}--\eqref{eq:GTLR} is ${\cal O}(PM^{3})$, where $M=N_{\text{f}}N_{xy}$, with $N_{\text{f}}$ being sufficiently smaller than $N_z^{(\text{part})}$.
Consequently, further computational cost reduction for $N_z$ is achieved by making use of the advantages of the RGF method and SCG method within the RSFD formalism in comparison with the conventional procedure.\cite{PRB86-195406}

\begin{figure}
\includegraphics[width=8.0cm]{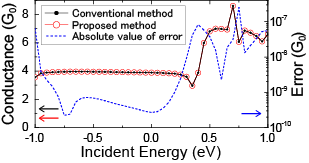}
\caption{\label{fig:Figure2}(Color online) conductance spectra of DWCNT (solid lines) and computational error of the proposed method (dashed line).}
\end{figure}

%%%
%%% %\section{Accuracy test \& Applications\label{sec:Applications}}
%%%
We evaluated the computational error between the conductance spectra using the Green's function matrices obtained by the proposed method and the conventional method\cite{PRB86-195406} for the BN-doped (4,4)@(8,8) DWCNT to verify that the proposed method does not deteriorate the computational accuracy.
The unit cell contains 192 atoms, and boron and nitrogen atoms are substitutionally co-doped into the outer tube along the circumferential direction. The dimensions are $L_{x(y)}=19.5~{\text{\AA}}$ and $L_{z}^{\text{(part)}}=9.84~{\text{\AA}}$.
For this verification, a supercell has twice the dimension in the $z$ direction that of the unit cell, that is, $L_z^{\text{(ext)}}= 2L_z^{\text{(part)}}$.
Using the electronic structure calculation code {\sc rspace}\cite{KikujiHirose2005,PRB72-085115} based on RSFD formalism within the DFT framework, the effective potential of the unit cell is calculated with $1\times 1\times 10$ $k$-points, $N_{xy}=80 \times 80$, $N_z^{(\text{part})}=40$, $N_{\text{f}}=4$, the NCPPs,\cite{ComputMaterSci14-000072,PRB43-001993,*PRL48-001425} and the local-density approximation~(LDA).\cite{CanJPhys58-001200} The Green's function calculation is collectively performed using the SCG method with $N_{E}=41$.
Figure~\ref{fig:Figure2} shows the conductance spectra of the supercell evaluated using the proposed method with $P = 2$ and the conventional method.
There are no notable errors more than $2.5 \times 10^{-7}~{\text{G}_0}$.
%%%
%%% Figure 2
%%%
%%%
%%%
%%%

Finally, we compared the computational times in calculating the Green's function submatrices for several systems using the proposed method with those using the conventional method.
Here the computational times are measured for systems in which (001) Si bulk is doped with B atoms.
We consider a primitive unit cell of (001) Si bulk, which contains four Si atoms and has the dimensions of $L_{x(y)}^{\text{(part)}}=3.84~{\text{\AA}}$ and $L_{z}^{\text{(part)}}=5.43~{\text{\AA}}$.
We measured the CPU time for calculating the submatrices of the Green's function matrix in a supercell with $L_{z}^{\text{(ext)}}=PL_{z}^{\text{(part)}}$, in which the $P$ unit cells are arranged along the $z$ direction, ($P=1,2,4$ and 8), and any of the four Si atoms in each unit cell is randomly replaced with a B atom. 
The effective potential of the unit cell is calculated using the {\sc rspace} code with $8 \times 8 \times 12$ $k$-points, $N_{xy}=12 \times 12$, $N_z^{(\text{part})}=16$, $N_{\text{f}}=4$, the NCPPs, and the LDA. 
We prepared small parts with $L_{x(y)}=n_{x(y)} \times L_{x(y)}^{\text{(part)}}$ by duplicating the unit cell $n_{x(y)}$ times in the $x(y)$ direction with $n_{x}=n_{y}=1,2$, and 4 to verify the influence of the cross-sectional size on the CPU time.
As summarized in Table~\ref{tab:ComputationalTime}, the CPU time for calculating the four submatrices of $\mathcal{G}_{1:P}$ with $N_E=16$ using the proposed method increases in proportion to $P$. We confirmed that the proposed method enabled us to notably reduce the computational cost and avoid computational difficulties in handling long systems.
Furthermore, one can see that the time ratio is not attenuated even in large cross-sectional systems. Since the proposed method has high weak-scaling efficiency for increasing $P$, the benefit of this method becomes more remarkable as the system size increases.

\begin{table}[tb]
\caption{
CPU time to obtain four submatrices of the Green's function matrix for B-doped (001)~Si bulk models.
The time ratio is evaluated as [calculating time + combining time]/[calculating time for $P=1$].
The calculations were performed on Intel\textregistered~Xeon\textregistered~CPU E5-2680.
}\label{tab:ComputationalTime}
\begin{tabular}{ccrrr} \hline
  		&		& \ \ calculating	& \ \ combining	& \ \ time \\
$(n_x,n_y)$	&\ $P$	& \ \ time (sec.)	& \ \ time (sec.)	& \ \ ratio \\ \hline \hline
(1,1)	 	& 1		& 175			& --			& 1.00	\\
	 	& 2		& 352			& 19 			& 2.12	\\
		& 4		& 669			& 56 			& 4.15	\\
  	 	& 8		& 1,347		& 130	 		& 8.45	\\ \hline
(2,2)	 	& 1		& 13,381		& --			& 1.00	\\ 
  	 	& 2		& 26,773		& 1,457 		& 2.11	\\ 
  	 	& 4		& 49,171		& 4,372 		& 4.00	\\
 	 	& 8		& 98,240		& 10,206 		& 8.10	\\ \hline
(4,4)	 	& 1		& 592,191		& --			& 1.00	\\
  	 	& 2		& 1,183,461		& 65,451 		& 2.11	\\
  	 	& 4		& 2,159,101		& 196,408 		& 3.98	\\
  		& 8		& 4,317,207		& 458,096 		& 8.06	\\ \hline
\end{tabular}
\end{table}

%%%
%%% Figure 3
%%%
\begin{figure}
\includegraphics[width=8.0cm]{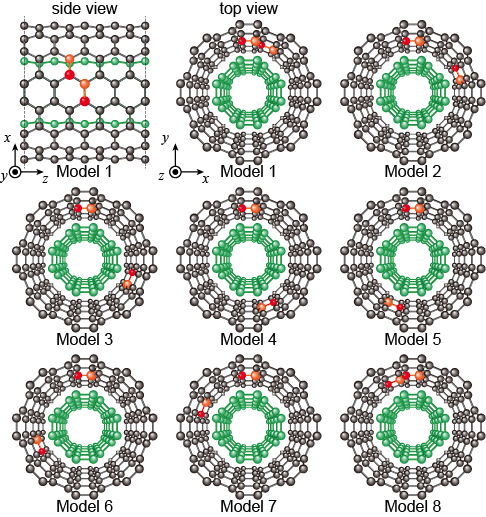}
\caption{\label{fig:Figure3}(Color online) Schematic view of the DWCNT unit-cell models. 
The grey and green spheres represent C atoms of the outer and inner carbon nanotubes, respectively, and the red and orange spheres represent N and B atoms, respectively.
Dashed lines represent the unit cell boundaries.}
\end{figure}
%%%
%%%
%%%

With the above verifications, it is confirmed that electron-transport properties can be estimated efficiently without the deterioration of computational accuracy using the proposed method.
As a practical demonstration of the method, we investigated the electron-transport properties of DWCNTs depending on the tube length.
As illustrated in Fig.~\ref{fig:Figure3}, eight DWCNT unit-cell models with different relative positions of the two BN dimers are prepared.
Here, the transport properties are evaluated for a model in which the same unit cells (Model~4) are periodically combined (periodic model) and in which different unit cells (Model~1--8) are randomly arranged (random model).
For the periodic model, the transport calculations are performed for the DWCNTs with $P=1$--$1,024$ to investigate the changes in the conductance spectrum according to the number of unit cells $P$ in the transition region. 
The numerical conditions and procedures to obtain the effective potential of each unit cell are the same as those mentioned above. Here, we ignore the electron-phonon couplings and evaluate the transport properties described in the ballistic regime.\cite{PRB82-085435}

\begin{figure}[tb]
\begin{center}
\includegraphics[width=8.0cm]{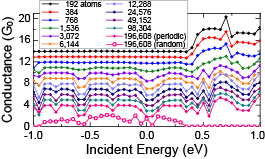}
\caption{\label{fig:Figure4}(Color online)
Conductance spectra for BN-doped DWCNTs. 
The filled-circle and open-circle plots represent the spectra of the periodic and random models, respectively.
The zero point of conductance is shifted by $1~{\text{G}_0}$ between the adjacent spectra, where the vertical scale represents the value for the 196,608-atom models~($P=1,024$). The incident energy is measured from the Fermi level.
}
\end{center}
\end{figure}

\begin{figure*}[tb]
\begin{center}
\includegraphics[width=16.0cm]{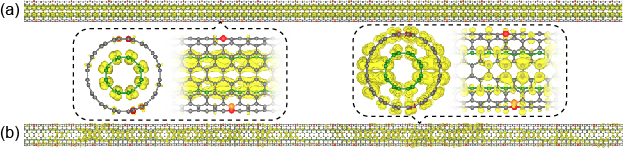}
\caption{\label{fig:Figure5}(Color online)
Distributions of the scattering wave functions in the conduction channels of the 6,144-atom model ($P=32$). Panels (a) and (b) illustrate the charge density of scattering wave functions with the highest transmission at the Fermi energy and $0.7$~eV above the Fermi energy, respectively. Insets surrounded by dashed lines depict cross-sectional and enlarged views. The key to the symbols is the same as that in Fig.~\ref{fig:Figure3}. The value of the yellow isosurface is $6.7 \times 10^{-4}$~electron/${\text{\AA}}^3$.
}
\end{center}
\end{figure*}

As shown in Fig.~\ref{fig:Figure4}, for systems with small $P$, the conductance quantization is observed at the low incident energy, where there are four conduction channels with a transmission probability of almost unity.
In the low energy region, the incident electrons flow through the intra-states of the inner and outer tubes, as depicted in Fig.~\ref{fig:Figure5}(a). 
Therefore, there are high-transmission channels due to the inner-tube states that are insensitive to the scattering potential created by the BN dimers in the outer tube.
However, the channel transmissions are notably reduced at the high incident energy because of the inter-states between the inner and outer tubes, as illustrated in Fig.~\ref{fig:Figure5}(b). 
As $P$ increases, dips appear in the conductance spectra even at the low incident energy, and they gradually become prominent.
This reflects the energy gap in the energy band structure that appears in an infinite DWCNT with a periodic scattering potential, as in the Kronig-Penney model.
On the contrary, for the random model with $P= 1,024$ (196,608 atoms, $\sim{1.0}~{\text{$\mu${m}}}$), overall conductance is suppressed as plotted in Fig.~\ref{fig:Figure4}.
Although the transmission in most conduction channels is notably reduced due to the random scattering potential component widely distributed in DWCNT, there are still channels with high transmission to which the intra-states of the inner tube contribute.
Here, the random model with $P=1,024$ is prepared by randomly combining four different random models with $P=64$ and 96.
We also performed calculations using random models with different configurations, but the results are essentially the same. 
Performing the large-scale electron-transport calculation, we can discover this difference in the conductance spectra between the randomly and periodically doped models.

%%%%%%%%
%%%%%%%% Figure 4 Conductance Spectra
%%%%%%%%
%\begin{figure}[p]

%%%%%%%%
%%%%%%%% Figure 5 scattering wave function
%%%%%%%%
%\begin{figure}[p]

%%%
%%% %\section{Conclusion\label{sec:Conclusion}}
%%%
In summary, we developed an efficient first-principles algorithm for evaluating the electron-transport properties of long systems consisting of a vast number of atoms.
The Green's function of the whole transition region extended towards the transport direction was obtained by recursively combining the Green's functions of the adjoining parts one by one. 
The computational cost for calculating the submatrices of the Green's functions required to estimate the transport properties is proportional to the number of the combined parts. 
As a practical application, we demonstrated the electron-transport calculations of BN-doped DWCNTs containing up to 196,608 atoms.
The proposed method develops our understanding of experimental studies on the research and development of devices using low-dimensional materials, as reported by Refs.~\onlinecite{NatureNanotechnol5-000858,NanoLett12-000758,Science355-000271,BiosensBioele137-000255,Small15-1903025}. 
Moreover, even in a submicron scale structure containing millions of atoms, it is possible to efficiently evaluate the electron-transport properties without any deterioration of computational accuracy.

%%%
%%% Acknowledgments
%%%
\begin{acknowledgments}
This work was partially supported by the financial support from MEXT as a social and scientific priority issue (creation of new functional devices and highperformance materials to support next-generation industries) to be tackled by using post-K computer and JSPS KAKENHI Grant Numbers~JP16H03865, JP18K04873.
S.T. acknowledges the financial support from Deutsche Forschungsgemeinschaft through the project numbers 389895192 and 277146847.
The numerical calculations were carried out using the system B and the system C of the Institute for Solid State Physics at the University of Tokyo, the Oakforest-PACS of the Center for Computational Sciences at University of Tsukuba, and the K computer provided by the RIKEN Advanced Institute for Computational Science through the HPCI System Research project (Project ID: hp190172).
\end{acknowledgments}

% Create the reference section using BibTeX:
\bibliography{manuscript.bib}

%\begin{thebibliography}{999}
%\bibitem{Comment1}
%\end{thebibliography}

\end{document}